\documentclass[a4paper,conference, anonymous]{IEEEtran}
\IEEEoverridecommandlockouts
\ifCLASSINFOpdf
\else
\fi
\usepackage{graphicx}
\usepackage{multirow}
\usepackage{booktabs}
\usepackage{mathtools}
\usepackage{amsfonts}
\usepackage{amssymb}
\usepackage{amsmath}
\usepackage{bbm}
\usepackage{hyperref}
\usepackage{color}
\usepackage[dvipsnames]{xcolor}
\begin{document}
\title{Multi-features based Semantic Augmentation Networks for Named Entity Recognition in \\ Threat Intelligence}

%\author{\IEEEauthorblockN{Peipei Liu,Hong Li}
%\IEEEauthorblockA{School of Cyber Security,\\ %University of Chinese Academy of %Sciences\\Institute of Information %Engineering, \\Chinese Academy of Sciences\\
%Beijing, China \\
%Email: liupeipei@iie.ac.cn,lihong@iie.ac.cn}
%\and
%\IEEEauthorblockN{Hong Li}
%\IEEEauthorblockA{School of Cyber Security,\\ University of Chinese Academy of Sciences\\Institute of Information Engineering, \\Chinese Academy of Sciences\\
%Beijing, China \\
%Email: lihong@iie.ac.cn}
%\and
%\IEEEauthorblockN{James Kirk\\ and Montgomery Scott}
%\IEEEauthorblockA{Starfleet Academy\\
%San Francisco, California 96678--2391\\


\author{\IEEEauthorblockN{Peipei Liu\textsuperscript{1,2},
 Hong Li\textsuperscript{1,2}$^{\ast}$ \thanks{*Corresponding author}, Zuoguang Wang\textsuperscript{1,2}, Jie Liu\textsuperscript{1,2}, Yimo Ren\textsuperscript{1,2}, Hongsong Zhu\textsuperscript{1,2}}
\IEEEauthorblockA{\textsuperscript{1}School of Cyber Security, University of Chinese Academy of Sciences, Beijing, China}
\IEEEauthorblockA{\textsuperscript{2}Institute of Information Engineering, Chinese Academy of Sciences, Beijing, China}
\IEEEauthorblockA{\{liupeipei, lihong, liujie1, renyimo, zhuhongsong\}@iie.ac.cn,\\ wangzuoguang16@mails.ucas.ac.cn}\vspace{-0.8cm}}


\maketitle

\begin{abstract}
Extracting cybersecurity entities such as attackers and vulnerabilities from unstructured network texts is an important part of security analysis\iffalse since fast and accurate extraction technique can help researchers improve their working efficiency \fi. However, the sparsity of intelligence data resulted from the higher frequency variations and the randomness of cybersecurity entity names makes it difficult for current methods to perform well in extracting security-related concepts and entities. To this end, we propose a semantic augmentation method which incorporates different linguistic features to enrich the representation of input tokens to detect and classify the cybersecurity names over unstructured text. In particular, we encode and aggregate the constituent feature, morphological feature and part of speech feature for each input token to improve the robustness of the method. More than that, a token gets augmented semantic information from its most similar \textit{K} words in cybersecurity domain corpus where an attentive module is leveraged to weigh differences of the words, and from contextual clues based on a large-scale general field corpus. We have conducted experiments on the cybersecurity datasets \textit{DNRTI} and \textit{MalwareTextDB}, and the results demonstrate the effectiveness of the proposed method.
\end{abstract}

\begin{IEEEkeywords}
    cybersecurity, named entity recognition, multi-features, semantic augmentation, attention mechanism
\end{IEEEkeywords}
% \keywords{cybersecurity, named entity recognition, multi-features, semantic augmentation, attention mechanism}

% \fi
\IEEEpeerreviewmaketitle

\section{Introduction}
The cyber threat intelligence (CTI) is a collection of evidence-based information, which is often used to describe threat information for cyber assets\cite{4}. Named entity recognition (NER) in CTI aims to tag cybersecurity entity names with their corresponding types such as users, malicious programs, hackers and vulnerabilities from plenty of unstructured CTI text. Accurate and speedy NER can be beneficial for researchers to carry out security analysis and assessment as well as enhance the real-time and precision of network security situation awareness. Thus, NER in CTI plays a major role in supporting and achieving cybersecurity research. Researches about NER in CTI have been widely pursued in recent years, 
and they can be summarized in the following three categories: rules-based~\cite{ZhangSZ,Kushner,XLiao,63,66}, statistical characteristics-based~\cite{68,CRF3,SVM,CRF2,SVM1} and deep learning-based~\cite{IanPerera,VIEM,QinYa2,XurenWang,XBiLSTM-CRF,CASIE,ATT-CNN-BiLSTM,QinYa3}.

\begin{figure}[ht!]
\vspace{-0.1cm}
  \centering
      \setlength{\abovecaptionskip}{0.1cm}
  \includegraphics[width=1\linewidth]{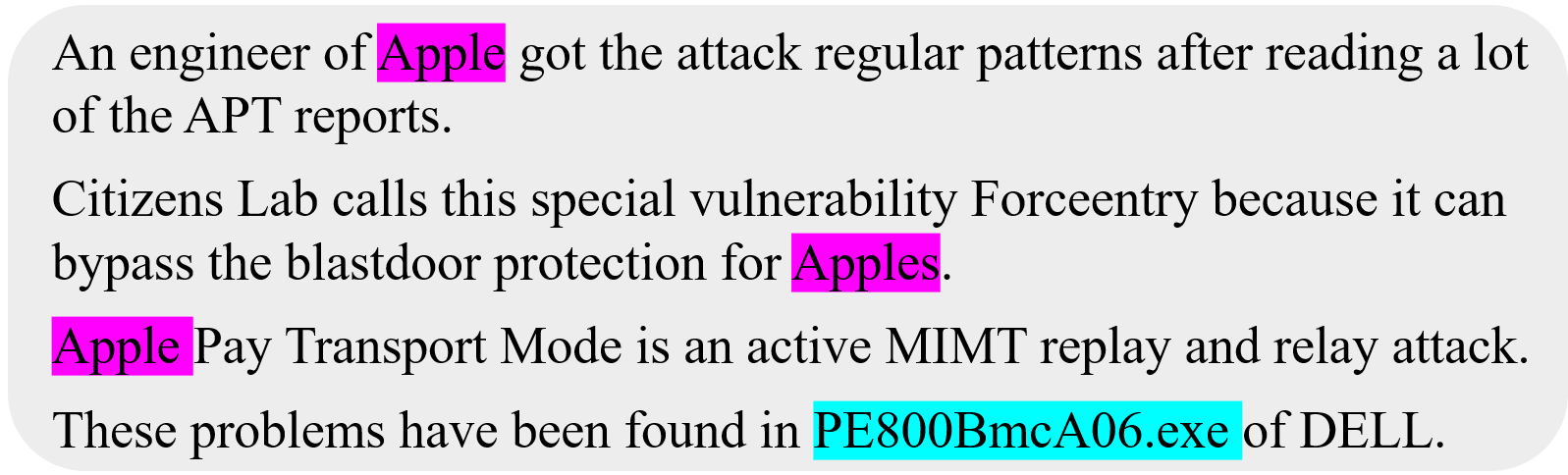}
  \caption{The sample sentences of CTI. What is the ‘Apple’? And How can we understand ‘PE800BmcA06.exe’?}
  \label{fig:one}
\end{figure}
\setlength{\textfloatsep}{3pt plus 1pt minus 1pt}

The rules-based methods use predefined rules and dictionaries to locate and extract entities with the advantages of accuracy, reliability and efficiency, while statistical characteristics-based methods apply feature engineering to learn representation for text and utilize machine learning algorithms to train models to recognize the entities. However, these approaches suffer from the limitations of poor portability and complex manual efforts.

Compared with the above two, deep learning-based methods have the unique advantages in the capability of representation learning and the semantic composition empowered by both the vector representation and neural processing\cite{DeepLearning}. Deep neural networks like recurrent neural networks(RNN)\cite{RNN1} and convolutional neural networks(CNN)\cite{CNN} with their variants learn intricate features and discover the semantic information automatically from raw data via non-linear activation functions in multiple processing layers\cite{NERSurvey}. These neural models can be trained in an end-to-end paradigm by gradient descent, and they have been widely used for NER in CTI such as~\cite{CASIE,XBiLSTM-CRF,ATT-CNN-BiLSTM,VIEM}.

Although the great success has been achieved by existing neural models, the sparsity of CTI data resulted from the high frequency variations and the randomness of cybersecurity entity names is not still given the priority. What's more, vague entity types such as \textit{‘APPLE’} for ORG and DEV\footnote{ORG means the \textit{organization}, and DEV means the \textit{device}. They are both the entity types in cybersecurity.} also make the existing methods inefficient and unperfect for recognition, as seen at Figure.\ref{fig:one}. To solve the problem, we design a multi-features based semantic augmentation model, which is inspired by the studies using semantic augmentation to improve model performance on other natural language processing (NLP) tasks~\cite{Kumar,Amjad,YuyangNie}. Our model consists of three main modules: \textit{Cybersecurity Domain Semantic Augmentation, General Domain Semantic Augmentation and Mixed Features Input}. \textit{Cybersecurity Domain Semantic Augmentation} enriches the representation of each input token from its most similar \textit{K} words in cybersecurity domain corpus, where an attentive module is leveraged to weigh differences of the words. In\textit{ General Domain Semantic Augmentation}, we refer to the embeddings produced by finetuning the pretrained BERT~\cite{BERT} as the external support for input words, as it can give sufficient contextual clues from the large-scale corpus and rich lexical knowledge. We regard both \textit{Cybersecurity Domain Semantic Augmentation} and \textit{General Domain Semantic Augmentation} as \textit{Semantic Augmentation (SA)}. As for \textit{Mixed Features Input}, except for common word embedding, morphological features encoded by CNN on character-level, part of speech features and component features are incorporated into the final representation of initial input words, before feeding into context encoding layers. After encodings of those three modules, a gate module is deployed on top of them to compute the most effective feature for entity detection and classification, with a followed softmax as tag decoder.

In summary, the contributions of this work are:
\begin{itemize}
\item We propose a multi-features based semantic augmentation model, aiming to detect and classify the cybersecurity entities effectively by solving the data sparsity and vague entity types problem.
\item The model is highly modularized, and it can be easily transferred for other related works such as event extraction, but not only NER in CTI.
\item We evaluate our approach on the CTI datasets \textit{DNRTI} and \textit{MalwareTextDB}, and the results show that our model is competitive with the current state-of-the-art methods. 
\end{itemize}

\section{Related Works}
As a specific aspect of the cybersecurity research, NER in CTI has experienced three stages of development. Most of the early works use hand-crafted rules to achieve decent performance. \cite{ZhangSZ} analyzes the regular expression matching method, and uses it to identify the types of attacks. The method is later used in Snort, l7-filter and Bro for the Deep Packet Inspection. A technique combining regular expression and ontology is then presented by \cite{Kushner} to extract entities from log files. The technique can use the formats of semi-structured files as features for type recognition and generating regular expressions, however, it cannot be applied to the extraction of unstructured files. \cite{XLiao} designs the iACE, an innovative solution for extraction of Indicators of Compromise (IOC) (e.g., malware signatures, botnet IPs) from public sources (e.g., blogs, forums, etc.) based on the combination of regular expression and syntax tree similarity. Furthermore, Bootstrapping algorithm is often viewed as the supplementary to improve the applicability and efficiency of rule-based matching methods~\cite{63,66,68}. \iffalse Especially, ~\cite{68} introduces an improved Bootstrapping method to extract security entities from blogs, twitter and other texts, which can increase the recall rate and maintain a high accuracy. \fi

Rules-based models work well when rules are exhaustive. Unfortunately, they cost a lot of manual efforts and are difficult to adapt to new tasks or new texts. To overcome the drawbacks of rules-based models, supervised methods based on machine learning such as Support Vector Machine (SVM)~\cite{SVM,SVM1}, Hidden Markov Models (HMM)~\cite{HMM1} and Conditional Random Fields (CRF)~\cite{CRF2,CRF3,QinYa} are applied to automatically learn similar patterns from unstructured text data. 

In recent years, with the development of deep learning, attention has been transferred to neural network models for NER in CTI. In~\cite{IanPerera}, cybersecurity entities are extracted by a NER system, which is comprised of a residual multi-task CNN trained on OntoNotes 5 and the automatically labeled vulnerability corpus.~\cite{VIEM} creates the VIEM, which employs a joint model based on character-level bidirectional gated recurrent units (BiGRU) and word-level BiGRU to identify the entities. \iffalse The character-level BiGRU is used to perform text encoding for each word while another BiGRU network is responsible for assigning labels to words.\fi ~\cite{XurenWang} uses the stacked bidirectional Long Short-Term Memory (BiLSTM) network to process the preliminary character-level features, then the resulting features are concatenated with word-level vectors from the GLoVe~\cite{Pennington} model serving as the input for encoder BiLSTM. Finally, two Dense layers and the Softmax function are designed for detection and classification.~\cite{XBiLSTM-CRF,HyejinShin,WEIXiao,QinYa2} apply CNN and BiLSTM neural networks as the basic framework to extract features of the domain information for NER in CTI. The differences among them are just attention mechanisms such as self-attention or multi-head attention allocating the corresponding weight of the extracted token feature, and tag decoders such as softmax or CRF obtaining association information among tags. ~\cite{CASIE} presents CASIE, a system that extracts information about cybersecurity events from text. The CASIE combines BiLSTM with multiple different linguistic features, providing a suite of competitive information extraction models for cybersecurity. 

Although these models have achieved fine results, they ignore the facts that vague cybersecurity entity types problem and the randomness and high-frequency variability of cybersecurity entity names could cause the data sparsity. In this paper, we will propose a model to solve both problems.
\section{The Proposed Method}
\label{method_framework}
The task of NER is conventionally posited as a standard sequence labeling problem, where an input sequence $ X = x_1,x_2,...x_N $ with $ N $ words is annotated with its corresponding labels ${Y} = {y}_{1}, {y}_{2}, ... , {y}_{N}$. The goal is thus to learn a parametrized mapping $ f_\theta : X \rightarrow {Y}$ from input words to output labels. To achieve the goal,
we firstly enter the mixed feature input of each word into an encoder \iffalse which consists of a four-layers BiLSTM and an one-layer multi-head attention \fi to get the basic contextual representation. Then, cybersecurity domain semantic augmentation representation and general domain semantic augmentation representation are combined with the basic contextual representation through a gate mechanism to enhance the semantic feature of the word. Finally, the resulting representation is passed into the decoder containing a feed forward neural networks and softmax to get label.  

\begin{figure*}[!ht]
  \centering
  \includegraphics[height=2.7in,width=5.8in]{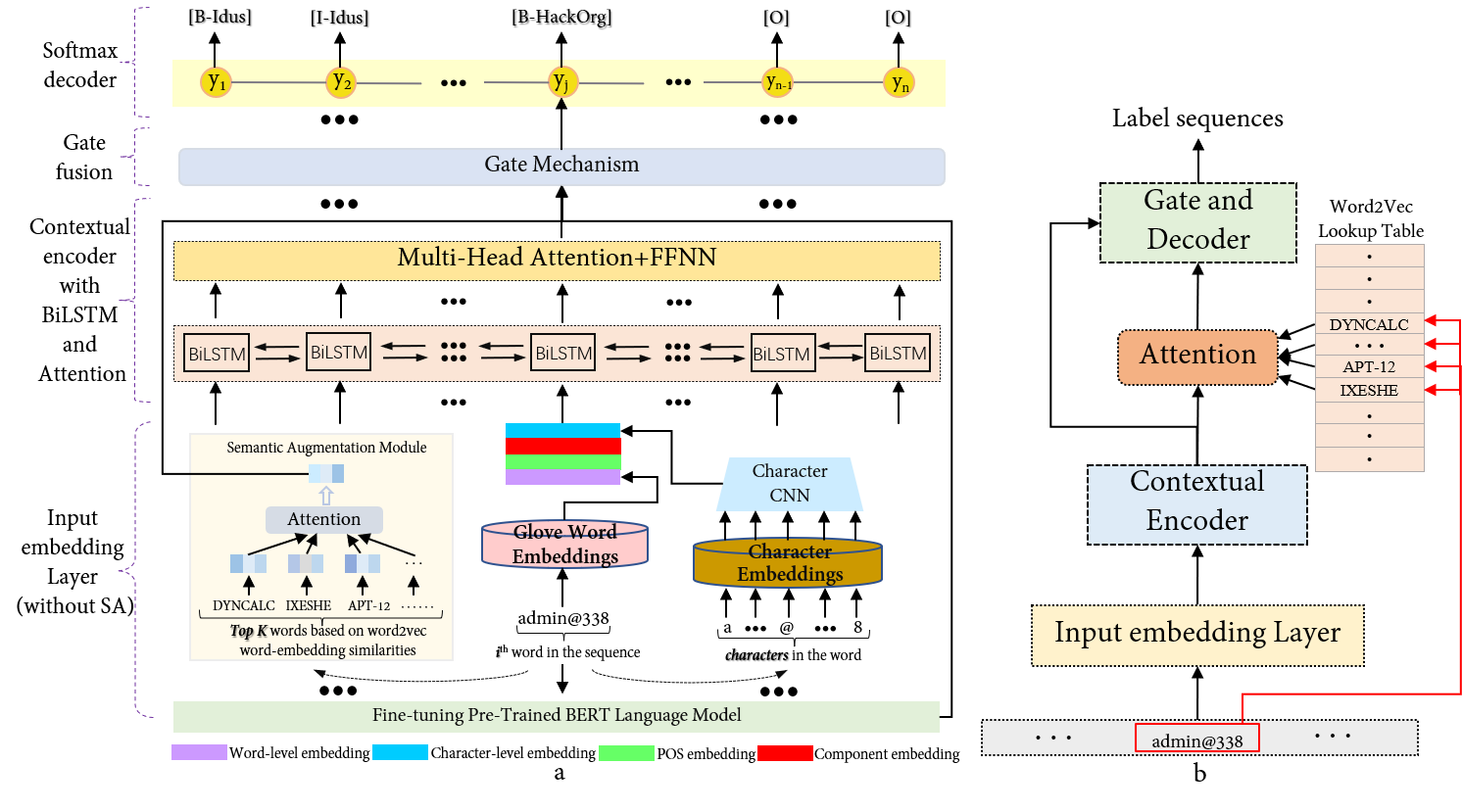}
    \setlength{\abovecaptionskip}{-0.3cm}
  \caption{Architecture of the proposed model: a is the whole framework with \textit{HSA}, while b denotes the diagram of detailed \textit{SSA} but ignores others.} \iffalse \centering{a is the whole framework of our method with \textit{HSA}, while b denotes the diagram of detailed \textit{SSA} but ignores others} \fi
  \label{fig:two}
  \vspace{-0.4cm}
\end{figure*}

In this section, we will describe main modules of the proposed neural network model one-by-one. Figure~\ref{fig:two} presents a high-level overview of the model framework architecture. 
\subsection{Basic model with Mixed Feature Input}
\label{MixedFeatureInput}

In this module,  we first construct the input representation for each word in the input sentence, and then pass it to the encoder to extract contextual feature.

\subsubsection{Mixed Feature Input}

Similar to the state-of-the-art NER approaches\cite{P-H,Aguilar,Jansson}, we also use word embeddings, character-level word embeddings and part-of-speech (POS) embeddings for NER in CTI. In addition, considering the characteristics of cybersecurity entity names, we take component embedding of each word into account. We will discuss each of these embeddings in the following text. 

Word embedding derived from the large corpus can learn co-occurrence statistics of words~\cite{Mikolov}. For each word, we retrieve one embedding from the lookup table initialized by Glove~\cite{Pennington}. For the out of vocabulary (OOV) words, embeddings are randomly initialized within $  (-\sqrt{\frac{3.0}{word_{dim}}},\sqrt{\frac{3.0}{word_{dim}}}) $, where $word_{dim}$ is the word dimension size. 

Character-level word embeddings have been proved useful in capturing morphological features for dealing with unseen words and OOV~\cite{CNN1}, which is very beneficial for NER in CTI. In our model, input character sequences are first passed through the embedding layer to get character embeddings, then a two-layers CNN with batch-norm is deployed to extract local features. Finally, CNN results are sent to a max-pooling layer with a Rectifier Linear Unit (ReLU)~\cite{relu} activation to map varying length vectors into fixed size representations of words.

POS can indicate the property of a word like noun or preposition, which is important for entity classification. For example, the POS tags of words \textit{‘Shamoon’} and \textit{‘StoneDrill}’ in the sentence \textit{‘Kaspersky believes both Shamoon and StoneDrill groups…’} are NN, indicating that they may be potential entities. Therefore, we utilize POS embeddings to enhance the representation ability of the input. In order to capture the contextual information of POS tags, we pretrain a CBOW~\cite{Mikolov} model based on POS tags with a context window of 3 tags and vector dimension size 30. In our work, the POS tag for each word is retrieved from Stanza~\cite{PengQi}.

Since the cybersecurity entity names contain various forms of characters, the component features may help us know the labels of words better. For example, words made up of numbers are more likely to be IPs or product models while words with all uppercase characters are more likely to be devices or vendors. Therefore, in this work, we get a separate one-hot lookup table to add the component feature with following options: \textit{allNum, allLower, allUpper, upperInit, mainNum, containNum, other}.

The above four types of embedding are then concatenated and linearly converted as the final embedding for each word. For the word $x_i$, we can get its embedding $w_i$ as:
{\setlength{\abovedisplayskip}{3pt}
\setlength{\belowdisplayskip}{3pt}
\begin{eqnarray}
w_i = [w_{wi},w_{mi},w_{pi},w_{ci}]W_E
\end{eqnarray}}
where $[\cdot,\cdot]$ is the concatenation operation with the same below, $W_{E}\in R^{98 \times 128}$ is a learnable matrix, $w_{wi}\in R^{50}$ denotes the word embedding, $w_{mi}\in R^{30}$ denotes the character-level word embedding, $w_{pi}\in R^{10}$ denotes the POS embedding, $w_{ci}\in R^{8}$ denotes the component feature.
\subsubsection{Encoder in the model}
For many sequence labeling tasks, it is beneficial to have access to both past and future contexts, and  \iffalse However, the original LSTM’s hidden state $h_i$ takes information only from past (where $0, 1, ..., i-1 $ ), knowing nothing about the future (where $i+1, ..., N$).\fi BiLSTM coincidentally is an elegant solution  \iffalse whose effectiveness has been proven by previous work~\cite{Dyer} \fi to achieve the goal. In our work, the mixed representation embedding of each word is passed through a two-layers BiLSTM to capture past and future information from different directions, and then the two separate hidden states are concatenated to form the final contextual representation. That is, for the ${i_{th}}$ word $x_i$ in the sentence, its contextual representation $h_i\in R^{256}$ from BiLSTM can be given as:
{\setlength{\abovedisplayskip}{3pt}
\setlength{\belowdisplayskip}{3pt}
\begin{eqnarray}
h_i= [LSTM_f(w_i), LSTM_b(w_i)]
\end{eqnarray}}where $LSTM_f$ denotes the forward LSTM and $LSTM_b$ denotes the backward LSTM.

The BiLSTM model can learn context features from input sequences automatically and effectively. However, these features have different contributions to NER. Fortunately, the multi-head attention can make the model learn relatively important features from different representation subspaces\iffalse by applying self-attention mechanism multiple times\fi~\cite{JianWang,JianWang1}. Hence, in our study, we capture relatively important features from outputs of the former BiLSTM by introducing the multi-head attention mechanism to benefit for NER. We firstly calculate the self-attention using the scaled dot-product attention function as:
{\setlength{\abovedisplayskip}{2pt}
\setlength{\belowdisplayskip}{2pt}
\begin{eqnarray}
Attention(Q,K,V)= softmax(\frac{QK^T}{\sqrt{d}})V
\end{eqnarray}}where $Q,K,V$ represent query, key, and value matrix, respectively. d is the dimension of $K$.

Then, the multi-head attention is computed by 8 parallel self-attention layers and the $t_{th}$ layer can be expressed by: 
{\setlength{\abovedisplayskip}{4pt}
\setlength{\belowdisplayskip}{4pt}
\begin{eqnarray}
head_t=Attention(HW_t^Q,HW_t^K,HW_t^V)
\end{eqnarray}}where $H \in R^{256} $ is the output of BiLSTM model and $W_t^Q,W_t^K,
W_t^V \in R^{256\times 32}$ are projection matrices. In result, the multi-head attention is the concatenation of $\{head_1,head_2,...,head_8\}$:
{\setlength{\abovedisplayskip}{4pt}
\setlength{\belowdisplayskip}{4pt}
\begin{eqnarray}
m=[head_1,...,head_t,...,head_8]W_M
\end{eqnarray}}where $W_M\in R^{256\times256}$ is the learnable parameter matrix.

After obtaining the word representation from the multi-head attention, we apply a feed-forward neural network (FFNN) to better aggregate and encode the features from different spaces.
{\setlength{\abovedisplayskip}{1pt}
\setlength{\belowdisplayskip}{1pt}
\begin{eqnarray}
m=FFNN(m)
\end{eqnarray}}

Finally, $m$ serves as the output from encoder and $m_i\in R^{256}$ acts for the feature representation of \iffalse $i_{th}$ word \fi $x_i$ in the input sentence.

\subsection{Cybersecurity Domain Semantic Augmentation}
\label{inter_aug}
We also call this as \textit{internal semantic augmentation} for the reason that the support corpora is from the experimental dataset. To handle the data sparsity, the \textit{internal semantic augmentation} captures commonalities of named entities in semantic space to improve the predicting effect of our model. Here, two internal semantic augmentation methods are designed named \textit{\bfseries Hard Semantic Augmentation(HSA)} and \textit{\bfseries Soft Semantic Augmentation(SSA)} respectively. We introduce details of these two methods in the following. 

Firstly, \iffalse let's start with an important auxiliary work, training \fi a word embedding model is trained on the unlabeled experimental dataset. For each word $x_i$ in the input sentence, we can get its embedding $v_i$ via the trained model. Subsequently, we can get the most similar \textit{K} words $\Phi_i=\{ x_{i1},...,x_{iK}\}$ of $x_i$ with corresponding embeddings $\Psi_i=\{v_{i1},...,v_{iK}\}$ based on the semantic space similarities.

\subsubsection{Hard Semantic Augmentation}
In this section, we enhance the semantic representation of $x_i$ by its \textit{K} nearest neighbors without training, which is called \textit{hard}. Since not all the \textit{K} words make equal contribution to assisting label prediction of $x_i$, it is important to get the most effective information from  different words. Thus, an attentive module is leveraged to weigh differences of the words. For each word $x_{ij}$, it is assigned a weight as:
{\setlength{\abovedisplayskip}{3pt}
\setlength{\belowdisplayskip}{3pt}
\begin{eqnarray}
\label{hsa_att}
\alpha_{ij}=\frac{exp(cosine(v_i, v_{ij}))}{\sum_k^K{exp(cosine(v_i, v_{ik}))}}
\end{eqnarray}}Then, we can get the semantic augmentation representation $w^I_i$ of $x_i$ by computing the weighted sum of all $\alpha_{ij}$ with their corresponding embeddings $v_{ij}$:
{\setlength{\abovedisplayskip}{2pt}
\setlength{\belowdisplayskip}{2pt}
\begin{eqnarray}
\label{hsa_att2}
w^I_i = \sum^K_{j=1}{\alpha_{ij}v_{ij}}
\end{eqnarray}}

\subsubsection{Soft Semantic Augmentation}
Different from the \textit{Hard Semantic Augmentation}, the acquisition of \textit{Soft Semantic Augmentation} needs the training procedure. We compute the augmentation representation of $x_i$ by measuring $\Psi_i$ and the $m_i$ derived from the encoder instead of $v_i$.
The same as above, the augmentation ability of the word $x_{ij}$ can be weighted by:
{\setlength{\abovedisplayskip}{3pt}
\setlength{\belowdisplayskip}{3pt}
\begin{eqnarray}
\label{ssa_att}
\beta_{ij}=\frac{exp(m_iW_Iv_{ij})}{\sum_k^K{m_iW_Iv_{ik}}}
\end{eqnarray}}where $W_I\in R^{256\times256}$ is a learnable parameter matrix. Next, the weighted semantic representation can be derived from:
{\setlength{\abovedisplayskip}{1pt}
\setlength{\belowdisplayskip}{1pt}
\begin{eqnarray}
\label{ssa_att2}
w^I_i = \sum^K_{j=1}{\beta_{ij}v_{ij}}
\end{eqnarray}}

After the above processing, the module ensures that semantic augmentation representations are weighted based on contributions of different words, and the most effective information for NER is obtained.
\subsection{General Domain Semantic Augmentation}
BERT~\cite{BERT} is a powerful language representation model, \iffalse and it is pre-trained on a large general field corpus. BERT \fi and it can explicitly model the correlation of a pair of tokens which is very helpful for solving vague cybersecurity entity types problem. For instance, the DEV \textit{‘APLLE’} often cooccur with \textit{‘operation system’}, while the ORG \textit{‘APPLE’} often appear with \textit{‘Google’}. In this work, we incorporate BERT into NER tasks after the finetuning procedure. As the BERT is pre-trained on the general field corpus, we also call this module as external support.

We take the final hidden state of the token corresponding to the target word from the BERT output as the word representation. If there are more than one token referring to a word, we sum them, i.e., we can get the representation of \textit{'gpu'} by summing \textit{'gp'} and \textit{'\#\#u'}. Then, a fully connect layer for every target lemma is added, which is the same as the last layer of the multi-head attention module. For the word $x_i$ with $P$ subtokens $\{\mu_{i1},...,\mu_{iP}\}$, its general semantic augmentation representation can be created by:
{\setlength{\abovedisplayskip}{1pt}
\setlength{\belowdisplayskip}{1pt}
\begin{eqnarray}
w_i^E=fc(\sum^P_{j=1}{e_{ij}})
\end{eqnarray}}where $e_{ij}$ is the token representation of $\mu_{ij}$, and $fc(\cdot)$ denotes the full connection.
\subsection{Decoder}
The gating mechanism regulates how much of the message propagates to the next step~\cite{gate}, and this provides the model a way to control contributions from the three modules in different text environments. We define a two-layers gate to compute the final comprehensive representation of $x_i$\iffalse for the label prediction. Formally,\fi:
{\setlength{\abovedisplayskip}{3pt}
\setlength{\belowdisplayskip}{3pt}
\begin{eqnarray}
f_i=\sigma(W_G[\theta_i,\lambda_i])
\end{eqnarray}
\begin{eqnarray}
\epsilon_i=f_i\odot\theta_i+(\boldsymbol{1}-f_i)\odot\lambda_i
\end{eqnarray}}where $W_G\in R^{256\times512}$ are the trainable parameters, $\sigma(\cdot)$ is the sigmoid function, $\odot$ stands for element-wise multiplication, and $\boldsymbol{1}$ is a vector whose elements are all 1. What's more, $\epsilon_i^1$ yielded in the first gate layer where $[\theta_i,\lambda_i]=[m_i,w^I_i]$ is then used as the input for the second gate layer with $[\theta_i,\lambda_i]=[\epsilon_i^1,w^E_i]$ to output $\epsilon_i^2$. 

Next, the $\epsilon_i^2$ is passed through a final softmax layer for label classification:
{\setlength{\abovedisplayskip}{2pt}
\setlength{\belowdisplayskip}{2pt}
\begin{eqnarray}
o_i=softmax({\epsilon_i^2}W_O+b_O)
\end{eqnarray}}where $W_O\in R^{256\times n_{l}}$ is a
parametrized matrix, $b_O \in R^{n_{l}}$ is the bias and $n_{l}$ indicates the number of entity types.

To extract the optimal predicted entity type $\hat{y}_i$ of $x_i$, we select the type corresponding to the maximum probability:
{\setlength{\abovedisplayskip}{3pt}
\setlength{\belowdisplayskip}{3pt}
\begin{eqnarray}
\hat{y}_i=argmax(o_i)
\end{eqnarray}}

During the training, we optimize our model with a cross-entropy loss:
{\setlength{\abovedisplayskip}{1pt}
\setlength{\belowdisplayskip}{1pt}
\begin{eqnarray}
loss = -\sum^N_{i=1}{\sum^{n_{l}}_{j=1}y_{i}^jlog(o_i^j)}
\end{eqnarray}}where $y_{i}^j$ equal to $0$ or $1$ is a binary indicator indicating whether $x_i$ truly belongs to $j_{th}$ entity type\iffalse, $N$ is the length of input sentence, $n_{l}$ is the number of entity labels\fi.

\section{Experiments}
\label{experiments}
We conduct extensive experiments on two open cybersecurity datasets \iffalse \textit{DNRTI}~\cite{XurenWang} and \textit{MalwareTextDB}~\cite{MalwareDB}  to verify the effectiveness of our model, and we also analyze the model with several different studies.\fi and our model is trained end-to-end by forward and back propagation. The project source can be publicly available at \url{https://github.com/LiuPeiP-CS/NER4CTI}.

\subsection{Datasets and Evaluation}
\label{dataset}
\textbf{Datasets:} \textit{DNRTI}~\cite{XurenWang} is a large-scale dataset for NER in CTI, consisting of 6574 annotated sentences and 36412 entities. All the entities can be divided into 13 categories, including hacker-organization(\textit{HackOrg}), attack(\textit{OffAct}), sample-file(\textit{SamFile}), security-team(\textit{SecTeam}), tool(\textit{Tool}), time(\textit{Time}), purpose(\textit{Purp}), area(\textit{Area}), industry(\textit{Idus}), organization(\textit{Org}), way(\textit{W-ay}), loophole(\textit{Exp}), features(\textit{Features}). 

\textit{MalwareTextDB}~\cite{MalwareDB} is an annotated malware database, annotated from 39 APT reports with a total of 6819 sentences and 10983 entities. All the entity tokens in MalwareTextDB can be labelled by 3 types: Action, Entity, and Modifier. Compared with \textit{DNRTI}, \textit{MalwareTextDB} is inferior to \textit{DNRTI} in the number of category and the number of instances per category. Moreover, \textit{MalwareTextDB} is sparser than \textit{DNRTI} due to the more sentences but the less entities.
%Both the datasets are in BIO tag scheme, where the B-tag indicates the beginning of an entity span, I-tag the inside and all other words are assigned the label O.% 

To approximately match the training data and testing data distributions, follow~\cite{XurenWang}, we randomly select 70\% of the original text as the training set, 15\% as validation set, and 15\% as the testing set on both datasets.

\textbf{Evaluation:} Following suggestions in \cite{XurenWang}, we evaluate Precision (P), Recall (R), and F1 scores with micro-averaging and adopt the strict evaluation criterion. Specifically, a predicted entity is correct only if its type and boundaries are correct.

\subsection{Setting}
\label{setting}
We use the pre-trained uncased ${BERT_{base}}$\footnote{\url{https://huggingface.co/bert-base-uncased}} model for finetuning, since we find that the cased BERT model performs slightly worse than the uncased model in this task. During finetuning, we use the training dataset to find the best settings for our task. We keep the dropout rate at 0.5. In addition, we have the learning rate warmup with the warmup proportion 0.002, the weight decay with the rate 1e-5, and the learning rate decay with the rate 1e-5. We finetune the BERT model in 100 epochs with the batch size 32 and learning rate 5e-5. 

For the BiLSTM in our basic model, our training uses 128 hidden states with batch size of 64. To train the model, we minimize the cross-entropy loss of the softmax class probability in \textit{Eq.16}. The model hyperparameters are updated using back-propagation by the Adam optimizer~\cite{Kingma}. The learning rate is 1e-3 with weight decay 1e-5, and the minimum learning rate is set to 5e-5. The model is regularized with a locked dropout rate of 0.3. We use 50-dimentional pre-trained word embeddings from Glove~\cite{Pennington}, 30-dimentional random initialized character embedding, 10-dimensional POS embedding, and 8-dimnesional component embedding as described in \ref{MixedFeatureInput}. We train our model for 200 epochs, and report the result for the model performing best on the 
validation set of the open data set collection. % The sequences are all filled to the maximum length of the both datasets, respectively.

Besides, we train the word2vec\cite{Mikolov} word embedding model\footnote{We also have experiments on other models, such as fastText, but find no significant differences. } for \ref{inter_aug} with min\_count 2, size 256 and window 3 in CBOW mode by using the gensim tool\footnote{\url{https://radimrehurek.com/gensim/}}.

\subsection{Main Results and Analysis}
Experiments are conducted on the open data collection introduced in \ref{dataset}. Table.\ref{tab:MB} and Table.\ref{tab:DNRTI} exhibit our results in comparison to previously published results\iffalse while Table.~\ref{tab:MalwareDB_types} and Table.~\ref{tab:DNRTI_types} show performance details of each type of entity on our final model \fi. We have five findings below:
% 以下两个表格是我们模型与之前方法对比的主要结果
%\renewcommand\arraystretch{1.2} % setting the hight
\begin{table}[t]
		%\setlength{\abovecaptionskip}{0cm} 
		%\setlength{\belowcaptionskip}{-0.2cm}
  \iffalse \caption{Main results of our model on MalwareTextDB dataset compared with previously published methods.} \fi
   \caption{Main comparison results on MalwareTextDB dataset.}
  \centering
  \label{tab:MB}
  \begin{tabular}{lllll}\toprule
  %\begin{tabular}{cllll}\toprule
  %\hline
    \iffalse
    \multirow{2}{*}{${Methods}$} & \multicolumn{2}{l}{${Valid Dataset}$} &\multicolumn{2}{l}{${Test Dataset}$}  \\ %\cline{2-3}\cline{4-5}
    \cmidrule(r){2-3}\cmidrule(r){4-5}
     & \textit{ACC} & \textit{F1}& \textit{ACC} & \textit{F1} \\ \hline
     \fi
    Methods & ACC& P&R & F1 \\\hline
    CRF\cite{MalwareDB} & -&51.7&27.0& 35.2 \\
    NaiveBayes+CRF\cite{MalwareDB} & -&45.9&36.3 & 40.3 \\
    BiLSTM+CRF\cite{CNN1}           &  83.74  &  40.18  & 46.64 &  43.17  \\
    IDCNN+CRF\cite{icdnn} &  85.40  &  50.00  & 46.46 &  48.17 \\
    CNN+BiLSTM+CRF \cite{CNN1} &  84.52  &  47.10  & 48.21 &  47.65 \\
    LSTM+BiLSTM+CRF \cite{XurenWang} &  87.53  &  -  & - &  47.52 \\
    $BERT finetuning_{ours} $   &    87.40   &    52.53   &    \bfseries63.49    &    57.49 \\
    $base model_{ours}     $    &    84.55   &    46.51   &    51.18    &    48.73 \\
    
    $FinalModel_{ours}$&\bfseries87.99 &\bfseries58.92 &62.01 & \bfseries60.43 \\\bottomrule
    %\hline
  \end{tabular}
  \vspace{-0.3cm}
\end{table}
\begin{table}[t]
		%\setlength{\abovecaptionskip}{0cm} 
		%\setlength{\belowcaptionskip}{-0.2cm}
  \iffalse \caption{Main results of our model on DNRTI dataset compared with other methods.} \fi
  \caption{Main comparison results on DNRTI dataset.}
  \centering
  \label{tab:DNRTI}
  \begin{tabular}{lllll}\toprule
    \iffalse
    \multirow{2}{*}{${Methods}$} & \multicolumn{2}{c}{${Valid Dataset}$} &\multicolumn{2}{c}{${Test Dataset}$}  \\ \cmidrule(r){2-3}\cmidrule(r){4-5}
     & \textit{ACC} & \textit{F1}& \textit{ACC} & \textit{F1} \\ \hline
    \fi
    Methods &ACC&P&R & F1 \\\hline
    IDCNN+CRF\cite{icdnn} &  98.66  &  74.42  & 77.40 &  75.88 \\
    BiLSTM+CRF \cite{CNN1} &  98.58  &  74.07  & 75.98 &  75.01 \\
    LSTM+LSTM+CRF\cite{XurenWang} &89.53&-& - & 67.09 \\
    LSTM+BiLSTM+CRF\cite{XurenWang} &90.85&-& - & 71.29 \\
    CNN+BiLSTM+CRF \cite{CNN1} & \bfseries 98.69  &  76.20  & 76.07 &  76.14 \\
    $BERT finetuing_{ours}$&94.33 &80.55 &80.27 & 80.41 \\
    $base model_{ours} $&94.90&81.18& 81.77 & 81.47 \\
    $FinalModel_{ours}$&95.43 &\bfseries86.16 & \bfseries84.54& \bfseries85.34 \\\bottomrule
  \end{tabular}
\end{table}
% 第一条数据，我们全对，对方全错；第二条数据，对方部分有错；第三条数据都有错
% \usepackage{multirow}
% \usepackage{multirow}
\begin{center}
\begin{table*}[htbp]
\caption{}
\label{tab:case}
\centering
\scriptsize
\begin{tabular}{|c|l|l|}
\hline
\multirow{3}{*}{Case1}                      &  GroundTruth&  \begin{tabular}[c]{@{}l@{}} We have previously observed \textcolor{Magenta}{APT19[HackOrg]} \textcolor{Apricot}{steal data[OffAct]} from law and investment firms for \textcolor{Salmon}{competitive economic[Purp]} purposes .\end{tabular}   \\ \cline{2-3} 
 & CNN+BiLSTM+CRF &       APT19[HackOrg] $\surd$     \quad  steal data[OffAct] $\surd$   \quad   economic[Idus] $\times$                   \\ \cline{2-3} 
 & Our Model &        APT19[HackOrg] $\surd$    \quad  steal data[OffAct] $\surd$    \quad    competitive economic[Purp] $\surd$                \\ \hline
\multicolumn{1}{|l|}{\multirow{3}{*}{Case2}} & GroundTruth &   In some \textcolor{Apricot}{attacks[OffAct]}, \textcolor{Magenta}{Whitefly[HackOrg]} has used a second piece of \textcolor{Green}{custom malware[Tool]}, \textcolor{Green}{Trojan.Nibatad[Tool]} .                                                                                                                             \\ \cline{2-3} 
\multicolumn{1}{|l|}{}  & CNN+BiLSTM+CRF &         attacks[OffAct]  $\surd$   \quad    Trojan.Nibatad[Tool]   $\surd$                                                                                                              \\ \cline{2-3} 
\multicolumn{1}{|l|}{}  & Our Model &          attacks[OffAct] $\surd$ \quad Whitefly[HackOrg] $\surd$ \quad custom malware[Tool] $\surd$ \quad Trojan.Nibatad[Tool] $\surd$                                                   \\ \hline
\multirow{3}{*}{Case3}                        & GroundTruth &   It seems \textcolor{Purple}{Eset[SecTeam]} has discovered and published on a new malware module created by \textcolor{Magenta}{Turla[HackOrg]}.                                                                                                                                                     \\ \cline{2-3} 
                        & CNN+BiLSTM+CRF &             Turla[HackOrg]      $\surd$                                                                                                                                    \\ \cline{2-3} 
                        & Our Model &     Eset[HackOrg]    $\times$    \quad            Turla[HackOrg]  $\surd$                                                                                                                 \\ \hline
\end{tabular}
\vspace{-0.5cm}
\end{table*}
\end{center}
\begin{table}[t]
  \caption{Module comparisons on the two datasets.}
  \centering

  \label{tab:alation_comparison}
  \begin{tabular}{cllllll}\toprule
\multirow{2}{*}{Methods}       & \multicolumn{3}{c}{MalwareTextDB}                                                      & \multicolumn{3}{c}{DNRTI}                                                  \\\cmidrule(r){2-4}\cmidrule(r){5-7}
                            & \multicolumn{1}{c}{P} & \multicolumn{1}{c}{R} & \multicolumn{1}{c}{F1} &        \multicolumn{1}{c}{P} & \multicolumn{1}{c}{R} & \multicolumn{1}{c}{F1} \\ \hline
\multicolumn{1}{l}{base model} &    46.51   &    51.18    &    48.73 &    81.18   &    81.77    &    81.47                \\ \hline
\multicolumn{1}{l}{base+SSA}  &    50.80   &    49.87    &    50.33 &    84.01   &    82.54    &    83.27                \\ \hline
\multicolumn{1}{l}{base+HSA}   &    49.78   &    48.47    &    49.12&    83.31   &    82.92    &    83.12                 \\ \hline
\multicolumn{1}{l}{base+BERT} &  \bfseries  59.29   &    59.65    &    59.47 &    85.11   &    83.48    &    84.29                 \\ \hline
\multicolumn{1}{l}{base+BERT+SSA} &58.04 &\bfseries63.06 & \bfseries60.44 &\bfseries86.32 &\bfseries84.63 & \bfseries85.47                 \\ \hline
\multicolumn{1}{l}{base+BERT+HSA}  &58.92 &62.01 & 60.43    &86.16 &84.54 & 85.34  \\     \bottomrule           
\end{tabular}
\end{table}
% rf https://wenku.baidu.com/view/891fe1956adc5022aaea998fcc22bcd127ff425b.html
\begin{enumerate}
    \item Our model improves NER performance on every dataset and this improvement is particularly large on F1. 
    \item CNN+BiLSTM performs better than BiLSTM, indicating the importance of applying character-level features for NER. % represented by CNN play an important role in improving the performance of NER;% 通过对比CNN+BiLSTM与BiLSTM，我们可以发现CNN所代表的字符级特征对于性能的提升起到了重要作用；
    \item  Mixed Feature (i.e. CNN+POS+Components) plays an important role in improving the performance of NER by comparing the CNN+BiLSTM to our base model, which confirms our thought in \ref{MixedFeatureInput}.% 通过对比CNN+BiLSTM与我们的base实验，我们发现，我们所提出的多特征融合（即CNN+POS+Components）,对安全数据集上的NER识别也很有帮助。
    \item Benefiting from the large-scale external corpus and transfer learning support, Bert can effectively adapt to CTI corpus and achieve satisfactory results \iffalse alone \fi for NER.% comparable results通过其强大的外部语料支撑和良好地迁移性，能够有效地适应安全语料，独自对NER实现不错的效果.
    \item The effect of our final model on DNRTI is stronger than MalwareTextDB, because the sparser characteristic of MalwareTextDB limits the ability of the model. % 我们地模型在DNRTI数据集上的提升效果，这是因为MalwareDB数据集过于稀疏，导致模型的提升能力也受到限制。
    \iffalse \item In Table.~\ref{tab:DNRTI_types}, we observe that recall rates of some types of entities are very high, even 1 (such as “Purp” and “Exp”) , which indicates that the model has a high coverage \iffalse of correct prediction \fi for such entities. \fi%  在表X中，我们更详细具体地展示了每种类型实体的识别性能，可以看到有些实体的识别效果确实很好，召回率和精准率都很高，有的召回率达到了1，这说明至少模型对此类实体进行了高覆盖。
\end{enumerate}

\subsection{Module Study}

% 这里主要是对时间进行比较

\iffalse
\renewcommand\arraystretch{1.2} % setting the hight
\begin{table}[t]
  \caption{The comparisons of time cost on DNRTI and MalwareTextDB.}
  \centering
  \label{tab:time_comparison}
  \begin{tabular}{lllll}\toprule
    \multirow{2}{*}{${Methods}$} & \multicolumn{2}{c}{${DNRTI}$} &\multicolumn{2}{c}{${MalwareDB}$}  \\ \cmidrule(r){2-3}\cmidrule(r){4-5}
     & \textit{Train} & \textit{Test}& \textit{Train} & \textit{Test} \\ \hline
     
    \iffalse $Methods$ &\textit{DNRTI} & \textit{MalwareTextDB} \\\hline \fi
    $base+HSA$ &\textit{R} & \textit{F1}&\textit{R} & \textit{F1} \\
    $base+SSA$ &\textit{R} & \textit{F1}&\textit{R} & \textit{F1} \\
    $base+BERT+HSA$ &\textit{R} & \textit{F1}&\textit{R} & \textit{F1} \\
    $base+BERT+SSA$ &\textit{R} & \textit{F1}&\textit{R} & \textit{F1} \\
    \bottomrule
  \end{tabular}
\end{table}
\fi

To determine which modules are responsible for our models better, we conduct six different incremental comparison experiments, as seen at Table.~\ref{tab:alation_comparison}. \iffalse including base model, base model+HSA, base model+SSA, base model+BERT, base model+BERT+SSA and base model+BERT+HSA, \fi We choose the best of them as the final model.

For both datasets, we observe that \iffalse finetuning BERT can get the similar result with our base model. \fi both internal semantic augmentation \iffalse (i.e. HSA or SSA) \fi and external semantic augmentation \iffalse (i.e. BERT) \fi can improve the performance of base model\iffalse, and they play an important role for the label classification \fi. Compared to HSA, SSA can get the slightly better result might due to the effect of powerful contextual augmentation representation. \iffalse Not only that, we find the HSA is much faster than SSA during training and testing (Table.\ref{tab:time_comparison}) \fi However, we find that the average time cost of SSA is about 11 times than HSA during training and testing, since the attention of HSA is computed in advance (Eq.\ref{hsa_att},\ref{hsa_att2}) while the computation of SSA is posteriori (Eq.\ref{ssa_att},\ref{ssa_att2}) \iffalse that is because of the complex computation of SSA with backpropagation \fi. To decide the best model, we further incorporate the HSA and SSA into base model with BERT respectively, which are base+BERT+HSA and base+BERT+SSA. Comprehensively considering the performance and time cost, we choose base+BERT+HSA as our final model. The results of our final model demonstrate the importance of gate and integration of different semantic augmentation modules.

\iffalse We have the same analysis for experiments on MalwareTextDB. \fi Besides, different from the results on DNRTI, the performance of different combinations on MalwareTextDB is quite fluctuant. The BERT can have an important impact on the final model while internal semantic augmentations have a slight effect. That may be because the strong contextual encoding capability is more useful than internal semantic augmentations in an environment where data is too sparse. \iffalse Although the SSA can get almost the same result with HSA, we recommend HSA to use owing to the time and the Occam's Razor.\fi

\subsection{Case Study}
\iffalse To help understanding, we list a few examples from the CDR test set in Table 6.\fi
In order to verify the effectiveness of our model for data sparsity intuitively, we select three representative cases from the test set, and compare their prediction results of CNN+BiLSTM+CRF with our model. Table.~\ref{tab:case} shows their prediction results. Next, we will analyze each case in detail. 

From Case 1, we find that sufficient contextual reasoning is necessary. Predicting the entity ‘competitive economic’ depends on the context words ‘for’ and ‘purpose’, and BERT based models deal with the problem better than pure BiLSTM.

Our model achieves more adequate results in Case 2 while the CNN+BiLSTM+CRF misses some predictions. Specifically, the similar words of ‘Whitefly’ and ‘custom malware’ are \{‘ESET’, ‘Symantec’, ‘Butterfly’, ‘LuckyMouse’\} and \{‘backdoor’, ‘tool’, ‘EternalBlue’, ‘Rocket’\} respectively, making us believe the effectiveness of semantic augmentation.

In Case 3, we detect all entities but CNN+BiLSTM+CRF still loses one entity. However, we unfortunately assign the wrong label to ‘Eset’. The reason may be that ‘Eset’ has the similar label space and semantic role to ‘Turla’.

\iffalse
\begin{table}[]
\begin{tabular}{|c|c|}
\hline
% color rf https://blog.csdn.net/weixin_29657201/article/details/78161537
GroundTruth                      
& \begin{tabular}[c]{@{}l@{}} {\textcolor{Magenta}{FireEye Labs{[}SecTeam{]}}} detects this phishing attack\\{[}Way{]}  and customers will be protected against the \\usage of  these sites in possible future campaigns{[}OffAct{]} . \end{tabular}\\ \hline

BiLSTM-CRF & \begin{tabular}[c]{@{}l@{}}FireEye Labs[SecTeam]  X  phishing attack[OffAct]    \\ campaigns{[}OffAct{]} $\surd$                                                                                      \end{tabular} \\ \hline
Our Model  & \begin{tabular}[c]{@{}l@{}} FireEye Labs[SecTeam]  X  phishing attack[OffAct]    \\ campaigns{[}OffAct{]} $\surd$                                                                                      \end{tabular} \\ \hline
\end{tabular}
\end{table}
\fi

\section{Conclusion}
In this paper, we seek to better leverage the augmentation semantics to solve the data sparsity for NER in CTI. We propose a new neural network model consisting of three parts: internal domain augmentation, external general augmentation and mixed Linguistic Features. The internal domain augmentation strengthens the semantics of input words by their most-K similar words in cybersecurity. We finetune the BERT model on cybersecurity datasets, and output representations of tokens are used for external general augmentation. What’s more, the POS feature, morphology feature and component feature construct the mixed Linguistic Features. The experiments on two CTI datasets show the effectiveness of our approach. \iffalse Except passing the word information into context encoder, the model is able to add some external knowledge to have better NER performance. \fi We expect that this idea can, at least partially, help perfect classification and detection tasks in cybersecurity research while data sparsity occurs.

\iffalse In a word, we find out that naive sequential labeling model has difficulty dealing with the data sparsity while relieving it is one of the strengths of our model, and generating more features to enhance semantic representation would be one possible direction for future research.\fi

\section{Acknowledgments}
This work was supported by the National Key R\&D Program of China (No.2018YFB0803402) and the National Natural Science Foundation of China (No.61842202). We thank our anonymous reviewers for their valuable suggestions. We also appreciate Beijing Key Laboratory of IOT Information Security Technology for providing the environment of our experiments.

\section{supplementary material}
In this supplementary material, we give more performance details of each entity type predicted by our model (Table.\ref{tab:MalwareDB_types} and Table.\ref{tab:DNRTI_types}). In Table.~\ref{tab:DNRTI_types}, we observe that recall rates of some types of entities are very high, even 1 (such as “Purp” and “Exp”) , which indicates that the model has a high coverage \iffalse of correct prediction \fi for such entities.

% 以下部分主要是实体类型的性能细节
\renewcommand\arraystretch{1.2} % setting the hight
\begin{table}[t]
		\setlength{\abovecaptionskip}{0cm} 
		\setlength{\belowcaptionskip}{-0.2cm}
  \caption{The performance details \iffalse of each type of entity in \fi on MalwareTextDB dataset.}
  \centering
  \label{tab:MalwareDB_types}
  \begin{tabular}{llll}\toprule
  %\begin{tabular}{cllll}\toprule
    Entity\ Types &P&R & F1 \\\hline
    Action           &    59.89   &    67.61    &   63.52 \\
    Entity            &    59.76   &    60.12    &    59.94 \\
    Modifier            &    54.21   &    58.86    &    56.44 \\
    \bottomrule
  \end{tabular}
\end{table}

\renewcommand\arraystretch{1.2} % setting the hight
\begin{table}[t]
		\setlength{\abovecaptionskip}{0cm} 
		\setlength{\belowcaptionskip}{-0.2cm}
  \caption{The performance details \iffalse of each type of entity in \fi on DNRTI dataset.}
  \centering
  \label{tab:DNRTI_types}
  \begin{tabular}{llll}\toprule
  %\begin{tabular}{cllll}\toprule
    Entity \ Types &P&R & F1 \\\hline
    Area           &    86.26 & 84.26&  85.26 \\
    Exp            &    98.51 & 100&  99.25 \\
    Features            &    91.27  &99.14&  95.04 \\
    HackOrg            &    82.92&  81.79 & 82.35 \\
    Idus            &    91.04 & 94.57  &92.78 \\
    OffAct            &   85.94  &73.33 & 79.14  \\
    Org           &   72.06&  71.53 & 71.79 \\
    Purp            &    85.82 & 100 & 92.37 \\
    SamFile            &    96.38  &85.89&  90.83 \\
    SecTeam            &    88.36&  84.87 & 86.58  \\
    Time            &    87.65 & 88.17  &87.91  \\
    Tool            &    78.85 & 70.51&  74.45 \\
    Way            &    84.21  &97.96 & 90.57  \\
    \bottomrule
  \end{tabular}
\end{table}


%\section*{Acknowledgment}




%\bibliographystyle{IEEEtran}
%\begin{thebibliography}{1}

\newpage
%\end{thebibliography}
\bibliographystyle{IEEEtran}
\bibliography{sample}

\begin{thebibliography}{10}
\providecommand{\url}[1]{#1}
\csname url@samestyle\endcsname
\providecommand{\newblock}{\relax}
\providecommand{\bibinfo}[2]{#2}
\providecommand{\BIBentrySTDinterwordspacing}{\spaceskip=0pt\relax}
\providecommand{\BIBentryALTinterwordstretchfactor}{4}
\providecommand{\BIBentryALTinterwordspacing}{\spaceskip=\fontdimen2\font plus
\BIBentryALTinterwordstretchfactor\fontdimen3\font minus
  \fontdimen4\font\relax}
\providecommand{\BIBforeignlanguage}[2]{{%
\expandafter\ifx\csname l@#1\endcsname\relax
\typeout{** WARNING: IEEEtran.bst: No hyphenation pattern has been}%
\typeout{** loaded for the language `#1'. Using the pattern for}%
\typeout{** the default language instead.}%
\else
\language=\csname l@#1\endcsname
\fi
#2}}
\providecommand{\BIBdecl}{\relax}
\BIBdecl

\bibitem{4}
J.~Williams, ``Cyber threat intelligence,'' \emph{SC magazine}, vol.~29, no.
  3SUPPL., pp. 62--63, 2018.

\bibitem{ZhangSZ}
ZHANG, Shu-Zhuang, LUO, Hao, FANG, and Bin-Xing, ``Regular expressions matching
  for network security,'' \emph{Journal of Software}, 2011.

\bibitem{Kushner}
M.~Balduccini, S.~Kushner, and J.~Speck, ``Ontology-driven data semantics
  discovery for cyber-security,'' in \emph{International Symposium on Practical
  Aspects of Declarative Languages}, 2015.

\bibitem{XLiao}
X.~Liao, K.~Yuan, X.~F. Wang, Z.~Li, and R.~Beyah, ``Acing the ioc game: Toward
  automatic discovery and analysis of open-source cyber threat intelligence,''
  in \emph{Acm Sigsac Conference on Computer \& Communications Security}, 2016.

\bibitem{63}
E.~M. Thelen, ``A bootstrapping method for learning semantic lexicons using
  extraction pattern contexts,'' 2002.

\bibitem{66}
C.~L. Jones, R.~A. Bridges, K.~M.~T. Huffer, and J.~R. Goodall, ``Towards a
  relation extraction framework for cyber-security concepts,'' \emph{ACM},
  2015.

\bibitem{68}
N.~Mcneil, R.~A. Bridges, M.~Iannacone, B.~Czejdo, N.~Perez, and J.~R. Goodall,
  ``Pace: Pattern accurate computationally efficient bootstrapping for timely
  discovery of cyber-security concepts,'' \emph{IEEE}, 2014.

\bibitem{CRF3}
R.~Lal, ``Information extraction of cyber security related terms and concepts
  from unstructured text,'' \emph{Dissertations \& Theses - Gradworks}, 2013.

\bibitem{SVM}
V.~MulwaD, W.~Li, A.~Joshi, T.~Finin, and K.~Viswanathan, ``Extracting
  information about security vulnerabilities from web text,'' in
  \emph{IEEE/WIC/ACM International Conference on Web Intelligence \&
  Intelligent Agent Technology}, 2011.

\bibitem{CRF2}
A.~Joshi, R.~Lal, T.~Finin, and A.~Joshi, ``Extracting cybersecurity related
  linked data from text,'' in \emph{Semantic Computing (ICSC), 2013 IEEE
  Seventh International Conference on}, 2013.

\bibitem{SVM1}
S.~More, M.~Matthews, A.~Joshi, and T.~Finin, \emph{A Knowledge-Based Approach
  to Intrusion Detection Modeling}.\hskip 1em plus 0.5em minus 0.4em\relax A
  Knowledge-Based Approach to Intrusion Detection Modeling, 2012.

\bibitem{IanPerera}
I.~Perera, J.~Hwang, K.~Bayas, B.~Dorr, and Y.~Wilks, ``Cyberattack prediction
  through public text analysis and mini-theories,'' in \emph{2018 IEEE
  International Conference on Big Data (Big Data)}, 2018.

\bibitem{VIEM}
\BIBentryALTinterwordspacing
Y.~Dong, W.~Guo, Y.~Chen, X.~Xing, Y.~Zhang, and G.~Wang, ``Towards the
  detection of inconsistencies in public security vulnerability reports,'' in
  \emph{28th {USENIX} Security Symposium ({USENIX} Security 19)}.\hskip 1em
  plus 0.5em minus 0.4em\relax Santa Clara, CA: {USENIX} Association, Aug.
  2019, pp. 869--885. [Online]. Available:
  \url{https://www.usenix.org/conference/usenixsecurity19/presentation/dong}
\BIBentrySTDinterwordspacing

\bibitem{QinYa2}
Y.~Qin, G.~Shen, W.~Zhao, and Y.~Chen, ``Research on the method of network
  security entity recognition based on deep neural network,'' \emph{Journal of
  Nanjing University(Natural Science)}, 2019.

\bibitem{XurenWang}
X.~Wang, X.~Liu, S.~Ao, N.~Li, and X.~Zhang, ``Dnrti: A large-scale dataset for
  named entity recognition in threat intelligence,'' in \emph{2020 IEEE 19th
  International Conference on Trust, Security and Privacy in Computing and
  Communications (TrustCom)}, 2020.

\bibitem{XBiLSTM-CRF}
Pingchuan, B.~Ma, Z.~Jiang, N.~Lu, Z.~Li, and Jiang, ``Cybersecurity named
  entity recognition using bidirectional long short-term memory with
  conditional random fields,'' \emph{Tsinghua Science and Technology}, vol.
  v.26, no.~03, pp. 11--17, 2021.

\bibitem{CASIE}
T.~Satyapanich, F.~Ferraro, and T.~Finin, ``Casie: Extracting cybersecurity
  event information from text,'' \emph{Proceedings of the AAAI Conference on
  Artificial Intelligence}, 2020.

\bibitem{ATT-CNN-BiLSTM}
F.~Ren, Z.~Jiang, and J.~Liu, ``Integrating an attention mechanism and deep
  neural network for detection of dga domain names,'' in \emph{2019 IEEE 31st
  International Conference on Tools with Artificial Intelligence (ICTAI)},
  2019.

\bibitem{QinYa3}
Y.~Qin, G.~W. Shen, W.~B. Zhao, Y.~P. Chen, Y.~U. Miao, and X.~Jin, ``A network
  security entity recognition method based on feature template and
  cnn-bilstm-crf,'' \emph{Frontiers of Information Technology and Electronic
  Engineering}, vol. 020, no. 006, pp. P.872--884, 2019.

\bibitem{DeepLearning}
Y.~Lecun, Y.~Bengio, and G.~Hinton, ``Deep learning,'' \emph{Nature}, vol. 521,
  no. 7553, p. 436, 2015.

\bibitem{RNN1}
G.~Toderici, S.~M. O'Malley, S.~J. Hwang, D.~Vincent, D.~Minnen, S.~Baluja,
  M.~Covell, and R.~Sukthankar, ``Recurrent neural network regularization.''

\bibitem{CNN}
Y.~Lecun, B.~Boser, J.~Denker, D.~Henderson, R.~Howard, W.~Hubbard, and
  L.~Jackel, ``Backpropagation applied to handwritten zip code recognition,''
  \emph{Neural Computation}, vol.~1, no.~4, pp. 541--551, 2014.

\bibitem{NERSurvey}
J.~Li, A.~Sun, J.~Han, and C.~Li, ``A survey on deep learning for named entity
  recognition,'' \emph{IEEE Transactions on Knowledge and Data Engineering},
  vol.~PP, no.~99, pp. 1--1, 2020.

\bibitem{Kumar}
V.~Kumar, H.~Glaude, C.~D. Lichy, and W.~Campbell, ``A closer look at feature
  space data augmentation for few-shot intent classification,'' 2019.

\bibitem{Amjad}
\BIBentryALTinterwordspacing
M.~Amjad, G.~Sidorov, and A.~Zhila, ``Data augmentation using machine
  translation for fake news detection in the {U}rdu language,'' in
  \emph{Proceedings of the 12th Language Resources and Evaluation
  Conference}.\hskip 1em plus 0.5em minus 0.4em\relax Marseille, France:
  European Language Resources Association, May 2020, pp. 2537--2542. [Online].
  Available: \url{https://aclanthology.org/2020.lrec-1.309}
\BIBentrySTDinterwordspacing

\bibitem{YuyangNie}
Y.~Nie, Y.~Tian, X.~Wan, Y.~Song, and B.~Dai, ``Named entity recognition for
  social media texts with semantic augmentation,'' 2020.

\bibitem{BERT}
J.~Devlin, M.~W. Chang, K.~Lee, and K.~Toutanova, ``Bert: Pre-training of deep
  bidirectional transformers for language understanding,'' 2018.

\bibitem{HMM1}
G.~Zhou and J.~Su, ``Named entity recognition using an hmm-based chunk
  tagger,'' in \emph{Proceedings of the 40th Annual 82 Meeting of the
  Association for Computational Linguistics, July 6-12, 2002, Philadelphia, PA,
  USA.}, 2002.

\bibitem{QinYa}
Y.~Qin, G.~Shen, and Y.~U. Hongxing, ``Large-scale network security entity
  recognition method based on hadoop,'' \emph{CAAI Transactions on Intelligent
  Systems}, 2019.

\bibitem{Pennington}
J.~Pennington, R.~Socher, and C.~Manning, ``Glove: Global vectors for word
  representation,'' in \emph{Conference on Empirical Methods in Natural
  Language Processing}, 2014.

\bibitem{HyejinShin}
H.~Shin, W.~Shim, J.~Moon, J.~W. Seo, S.~Lee, and Y.~H. Hwang, ``Cybersecurity
  event detection with new and re-emerging words,'' ser. ASIA CCS '20.\hskip
  1em plus 0.5em minus 0.4em\relax New York, NY, USA: Association for Computing
  Machinery, 2020, p. 665–678.

\bibitem{WEIXiao}
C.~Y. WEI~Xiao, QIN~Yongbin, ``A network security named entity recognition
  method based on the component cnn,'' \emph{Computer \& Digital Engineering},
  vol. v.48;No.363, no.~01, pp. 111--116, 2020.

\bibitem{P-H}
P.-H. Li, R.-P. Dong, Y.-S. Wang, J.-C. Chou, and W.-Y. Ma, ``Leveraging
  linguistic structures for named entity recognition with bidirectional
  recursive neural networks,'' in \emph{Proceedings of the 2017 Conference on
  Empirical Methods in Natural Language Processing}.\hskip 1em plus 0.5em minus
  0.4em\relax Copenhagen, Denmark: Association for Computational Linguistics,
  Sep. 2017, pp. 2664--2669.

\bibitem{Aguilar}
G.~Aguilar, S.~Maharjan, A.~P. López-Monroy, and T.~Solorio, ``A multi-task
  approach for named entity recognition in social media data,'' 2019.

\bibitem{Jansson}
P.~Jansson and S.~Liu, ``Distributed representation, {LDA} topic modelling and
  deep learning for emerging named entity recognition from social media,'' in
  \emph{Proceedings of the 3rd Workshop on Noisy User-generated Text}.\hskip
  1em plus 0.5em minus 0.4em\relax Copenhagen, Denmark: Association for
  Computational Linguistics, Sep. 2017, pp. 154--159.

\bibitem{Mikolov}
T.~Mikolov, K.~Chen, G.~Corrado, and J.~Dean, ``Efficient estimation of word
  representations in vector space,'' \emph{Computer Science}, 2013.

\bibitem{CNN1}
X.~Ma and E.~Hovy, ``End-to-end sequence labeling via bi-directional
  lstm-cnns-crf,'' 2016.

\bibitem{relu}
X.~Glorot, A.~Bordes, and Y.~Bengio, ``Deep sparse rectifier neural networks,''
  vol.~15, 01 2010.

\bibitem{PengQi}
P.~Qi, Y.~Zhang, Y.~Zhang, J.~Bolton, and C.~D. Manning, ``Stanza: A python
  natural language processing toolkit for many human languages,'' 2020.

\bibitem{JianWang}
J.~Wang, X.~Chen, Y.~Zhang, Y.~Zhang, and X.~Wang, ``Document-level biomedical
  relation extraction using graph convolutional network and multi-head
  attention (preprint),'' 2019.

\bibitem{JianWang1}
A.~Vaswani, N.~Shazeer, N.~Parmar, J.~Uszkoreit, L.~Jones, A.~N. Gomez,
  L.~Kaiser, and I.~Polosukhin, ``Attention is all you need,'' \emph{arXiv},
  2017.

\bibitem{gate}
Y.~Luan, D.~Wadden, L.~He, A.~Shah, M.~Ostendorf, and H.~Hajishirzi, ``A
  general framework for information extraction using dynamic span graphs,'' in
  \emph{Proceedings of the 2019 Conference of the North}, 2019.

\bibitem{MalwareDB}
S.~K. Lim, A.~O. Muis, W.~Lu, and C.~H. Ong, ``{M}alware{T}ext{DB}: A database
  for annotated malware articles,'' in \emph{Proceedings of the 55th Annual
  Meeting of the Association for Computational Linguistics (Volume 1: Long
  Papers)}.\hskip 1em plus 0.5em minus 0.4em\relax Vancouver, Canada:
  Association for Computational Linguistics, Jul. 2017, pp. 1557--1567.

\bibitem{Kingma}
D.~Kingma and J.~Ba, ``Adam: A method for stochastic optimization,''
  \emph{Computer Science}, 2014.

\bibitem{icdnn}
F.~Yu and V.~Koltun, ``Multi-scale context aggregation by dilated
  convolutions,'' 11 2016.

\end{thebibliography}

\end{document}